\documentclass{iopart}

\usepackage{iopams}
\usepackage{graphicx}
\usepackage{a4wide}
\usepackage[colorlinks=true,linkcolor=blue,citecolor=blue,filecolor=cyan,urlcolor=blue,breaklinks=true]{hyperref}

\usepackage{cite}

\newcommand{\mrd}{\mathrm d}
\newcommand{\mre}{\mathrm e}
\newcommand{\pstat}{p^{\mathrm s}}
\newcommand{\mean}[1]{\left\langle #1 \right\rangle}
\newcommand{\eqref}[1]{(\ref{#1})}

\newcommand{\scl}{z}  

\begin{document}
\title{Extreme fluctuations of active Brownian motion}

\author{Patrick Pietzonka, Kevin Kleinbeck and Udo Seifert}

\address{II. Institut f\"ur Theoretische Physik, Universit\"at Stuttgart, 70550 Stuttgart, Germany}

\begin{abstract}
  In active Brownian motion, an internal propulsion mechanism interacts
  with translational and rotational thermal noise and other internal fluctuations to produce directed 
  motion. We derive the distribution of its extreme fluctuations and
  identify its universal properties using large deviation theory. 
  The limits of slow and fast internal dynamics give rise to a
  kink-like and parabolic behavior of the corresponding rate functions,
  respectively. For dipolar Janus particles in two and three dimensions 
  interacting with a field, we predict a novel symmetry akin to, but
  different from, the one related to entropy production. Measurements 
  of these extreme fluctuations could thus be used to infer properties 
  of the underlying, often hidden, network of states.\\[1\baselineskip]
\end{abstract}
\vspace{1cm}

\section{Introduction}
Stochastic thermal noise prevents predicting the dynamics of small systems
exactly.  In biological systems, one observes a persistent interplay between
this noise and the goal of the system to keep vitally important processes
running in a preferred direction. Such directed processes take place under
non-equilibrium conditions, which have to be maintained by a permanent supply
of energy. For instance, molecular motors burn chemical fuel molecules,
typically ATP, to displace cargo in a preferred direction \cite{astu97,howard}. On
larger length scales, many types of bacteria and some eucaryotic cells use
flagella to generate propulsion forces that allow directed swimming
\cite{cate12,gold15}.  Due to recent advances in nanotechnology, artificial
systems with a variety of propulsion mechanisms have been built. For instance,
Janus particles are colloidal particles with two differently coated halves,
which react differently to external heating or to chemical non-equilibrium
conditions and thereby generate self-phoretic motion
\cite{paxt04,gole05,jian10,butt12}. The superposition of the directed motion
in these systems with the ubiquitous Brownian motion is classified as
\textit{active} Brownian motion \cite{erdm00,hows07,gole09,ring10,ebbe10}, for which
phenomena like the interaction with confining boundaries
\cite{volp11,elge15,ghos13,das15} and external fields
\cite{pala10,encu11,zoet12,kian15} and emergent hydrodynamic and collective
properties like phase separation and swarming
\cite{rama10,gros12,wens12,marc13,pala13,butt13,spec14} have been intensely
investigated recently.

An important characteristic of active Brownian motion is the probability
distribution of fluctuations in the displacement of particles. As a
consequence of the central limit theorem, typical fluctuations during long
time intervals can be well described by a Gaussian distribution. In contrast,
extreme fluctuations are usually non-Gaussian and depend crucially on the
microscopic details of the underlying system.  In this paper, we provide a
full characterization of extreme fluctuations for active Brownian motion on
large time scales.  Earlier studies have analyzed the non-Gaussian behavior in
terms of the low order cumulants of the distribution for the particular case
of Janus particles \cite{hage09,hage11,sadj15,sevi15}.

As a mathematical tool, we make use of large deviation theory to calculate the
so-called rate function, which captures the exponential contributions to a
probability distribution \cite{touc09}. In stochastic thermodynamics, this
approach has led to the formulation of fluctuation theorems as a fundamental
property of extreme fluctuations \cite{evan93,lebo99,seif12}. These
fluctuation theorems, however, can only be applied when the system under
consideration does not contain any slow hidden degrees of freedom
\cite{mehl12}. If one is interested in the displacement of an active particle,
different internal states affecting the propulsion force must often
be considered as such a hidden degree of freedom.

So far, the only study of large deviations in the context of active Brownian
motion was conducted experimentally for an asymmetric particle interacting
with granular matter under non-equilibrium conditions, which revealed a
fluctuation theorem for the integrated velocity of the particle
\cite{kuma11,kuma15} as an instance of the isometric fluctuation relation
\cite{hurt11}. Recent experimental and numerical evidence suggests that
this fluctuation relation holds for phoretic self-propelled particles as well
\cite{fala16}. Yet, in these studies the velocity was measured in a frame
fixed to the particle. For active particles that undergo rotational diffusion,
as for example Janus particles, this integrated velocity differs from the
displacement perceived by an external observer. Our results are formulated in
the lab frame and can therefore be applied even if the rotational diffusion of
the particle cannot be observed or only with insufficient
precision.

The paper is organized as follows: In section~\ref{sec:discrete}, we present a
formalism using discrete internal states and biased diffusion for the
propulsion. To illustrate this model, molecular motors are presented as a toy
model before we analyze two analytically tractable limiting cases.  In
section~\ref{sec:janus}, we discuss the extreme fluctuations of the
displacement of a Janus particle with a dipole moment interacting with a
homogeneous field, for which we derive a novel symmetry similar to a
fluctuation theorem. In section~\ref{sec:generalization}, we generalize the
formalism from section~\ref{sec:discrete}, allowing for a more detailed
description of the propulsion mechanism. We conclude in
section~\ref{sec:summary}.

\section{Discrete internal states}
\label{sec:discrete}

\subsection{Langevin formalism}
\label{sec:discreteformalism}

As a simple but fairly general description of active Brownian motion, we
consider a system that is characterized by a continuous variable $x$ for the
displacement and a discrete set $\{i\}$ representing mesoscopic states, which
could be inaccessible to direct observation. The switching between these
states is modeled as a Markovian network with transition rates $w_{ij}$ from
state $i$ to state $j$. Assuming the system to be homogeneous along the
coordinate $x$, the rates $w_{ij}$ do not depend on $x$. The system performs
self-propelled motion along the coordinate $x$.  As essential ingredient of
the model, distinguishing active particles from merely externally pulled
passive particles, the self-propulsion force $f_i$ depends on the mesoscopic
state $i$.  At the present level of description, the microscopic mechanism
that actually generates the propulsion force in not yet included in the
model. As we will show below in Sec.~\ref{sec:generalization}, the states
$\{i\}$ and the forces $f_i$ emerge via coarse-graining from a more detailed
description.  The switching between the states $\{i\}$ will typically be
caused by thermal fluctuations with transition rates $w_{ij}$ obeying detailed
balance. In principle, however, it could involve an active component as well.

The Langevin equation for the displacement $x(t)$ at
time $t$ is
\begin{equation}
  \label{eq:langevin}
  \dot x(t)=\mu_i f_i+\zeta_i(t).
\end{equation}
For fixed $i$, this Langevin equation corresponds to the overdamped
dynamics of a particle with mobility $\mu_i$ driven by a force $f_i$. In
addition, the particle experiences thermal noise $\zeta_i(t)$ with the
statistical averages $\mean{\zeta_i(t)}=0$ and
$\mean{\zeta_i(t)\zeta_i(t')}=2D_i\delta(t-t')$. The diffusion coefficient
$D_i$ is related to the mobility via the Einstein relation $D_i=\mu_i
k_\mathrm{B} T$, with Boltzmann's constant $k_\mathrm{B}$ and the temperature
$T$ of the surrounding heat bath.

For a Janus particle, the variable $i$ can label different geometrical
alignments, while $x$ denotes its position in a one-dimensional channel (or
the projection of the position of a free Janus particle onto the $x$-axis).
If the self-propulsion mechanism generates no torque on the particle, the
internal transitions $w_{ij}$ describe the rotational diffusion of the
particle and satisfy detailed balance. For each state $i$ a different
component $f_i$ of the propulsion force acts in the direction of the
coordinate $x$. If the particle is not spherically symmetric, the mobility
$\mu_i$ can depend on the state $i$. For biological microswimmers, the
internal state can also encode different modes of motility, such as the
``run'' and the ``tumble'' mode in many types of bacteria.

The evolution of the joint probability
distribution of $x$ and $i$ is governed by a combination of a Fokker-Planck
equation for $x$ and a master equation for $i$,
\begin{equation}
  \partial_t p_i(x,t)=\left(-\mu_i
    f_i \partial_x+D_i\partial_x^2\right)p_i(x,t)
   +\sum_jL_{ij}p_j(x,t),
  \label{eq:fokkerplanck}
\end{equation}
where 
\begin{equation}
  \label{eq:markovmatrix}
  L_{ij}\equiv w_{ji}-r_i\delta_{ij}
\end{equation}
and $r_i\equiv\sum_\ell w_{i\ell}$ is the exit rate from
state $i$.  With this joint probability distribution, the distributions for
the internal variable $i$ and the position variable $x$ follow as
\begin{equation}
  p_i(t)=\int_{-\infty}^\infty\mrd x\,p_i(x,t)
\end{equation}
and
\begin{equation}
  p(x,t)=\sum_ip_i(x,t),
\end{equation}
respectively.
In the
long time limit, the distribution of the internal variable attains a stationary
distribution $\pstat_i$, which is independent of the initial distribution and
satisfies
\begin{equation}
  \label{eq:pstat}
  \sum_j L_{ij}\pstat_j=0.
\end{equation}

The ensemble average of the velocity of the system in the long time limit can be calculated
directly from the stationary distribution $\pstat_i$ as
\begin{equation}
  \label{eq:velocity}
  v\equiv\lim_{t\to\infty}\mean{x(t)/t}=\lim_{t\to\infty}\mean{\dot
    x(t)}=\sum_i\pstat_i\mu_i f_i.
\end{equation}
Along individual trajectories, the time averaged velocity $u\equiv x/t$ may differ
from the ensemble average $v$. However, the probability of such fluctuations
decreases exponentially with the length $t$ of the trajectory,
\begin{equation}
  p(x,t)\sim\mre^{-th(x/t)},
\end{equation}
characterized by of the so-called \textit{rate function} or \textit{large
  deviation function} \cite{touc09}
\begin{equation}
  \label{eq:ldf_def}
  h(u)\equiv-\lim_{t\to\infty}\frac{1}{t}\ln p(x=ut,t).
\end{equation}
Expanding $h(u)$ to second order around its minimum $h(v)=0$
yields a Gaussian approximation to the probability distribution, characterized by the effective diffusion coefficient
\begin{equation}
  \label{eq:Deff}
  D_\mathrm{eff}\equiv \lim_{t\to\infty}\frac{1}{2t}\left(\mean{x^2}-\mean{x}^2\right)\equiv\lim_{t\to\infty}C_2/(2t)=1/h''(v).
\end{equation}
Higher derivatives of $h(u)$ at $u=v$ relate to the higher order cumulants $C_n$.
The rate function captures the full range of non-Gaussian fluctuations in the
long-time limit and is often easier to interpret than a large set of
cumulants.

The standard approach to calculating the large deviation function
\cite{lebo99,touc09} starts with the transformation
\begin{equation}
  g_i(\lambda,t)\equiv\int_{-\infty}^\infty\mrd x\,\mre^{\lambda x}\,p_i(x,t)
\end{equation}
with a real variable $\lambda$. The Gaussian nature of the noise ensures that
tails of $p_i(x,t)$ decay like a Gaussian, therefore the integral converges
for all $\lambda$.  The evolution equation \eqref{eq:fokkerplanck} then
transforms to
\begin{equation}
  \label{eq:genfuncev}
  \partial_t g_i(\lambda,t)=\sum_j\mathcal{L}_{ij}(\lambda)\,g_j(\lambda,t)
\end{equation}
with the matrix
\begin{equation}
  \label{eq:L_general}
  \mathcal{L}_{ij}(\lambda)\equiv L_{ij}+(\mu_i f_i\lambda+D_i\lambda^2)\delta_{ij}.
\end{equation}
Eq.~\eqref{eq:genfuncev} can be solved using of an expansion in
eigenvectors of the matrix $\mathcal{L}_{ij}(\lambda)$. In the long-time
limit, the cumulant generating function, defined as
\begin{equation}
  \label{eq:CGF}
  \alpha(\lambda)\equiv\lim_{t\to\infty}\frac{1}{t}\ln g_i(\lambda,t)=\max_j\mathrm{Re}[\alpha_j(\lambda)],
\end{equation}
depends neither on $i$ nor on the initial distribution and is given by the
eigenvalue $\alpha_j(\lambda)$ with the largest real part. The
Perron-Frobenius theorem ensures that this eigenvalue has no imaginary part.
The cumulants $C_n$ of the distribution of $x$ in the long time limit follow
as
\begin{equation}
  \lim_{t\to\infty}C_n/t=\left.\partial_\lambda^n\alpha(\lambda)\right|_{\lambda=0}.
\end{equation}
 The cumulant generating function is convex, since it can be
written as the Legendre transform
\begin{equation}
  \label{eq:legendre1}
  \alpha(\lambda)=\max_u[\lambda u-h(u)]
\end{equation}
of the rate function $h(u)$. If the latter is convex as well,
$\alpha(\lambda)$ is differentiable for all $\lambda$ and the Legendre
transform can be inverted to
\begin{equation}
  \label{eq:legendre2}
  h(u)=\max_{\lambda}[\lambda u-\alpha(\lambda)].
\end{equation}

\subsection{Toy model: Two-state model for molecular motors}
\label{sec:casestudy}
\begin{figure}
  \centering
  \includegraphics[width=0.25\textwidth]{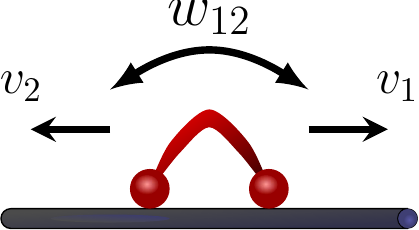}\qquad\qquad 
  \includegraphics[width=0.25\textwidth]{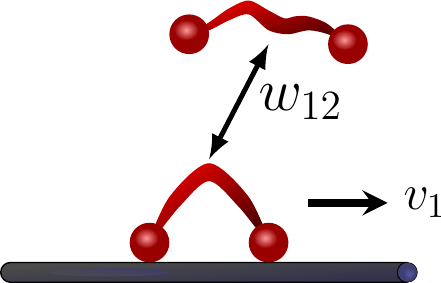}\\
  \includegraphics[width=0.4\textwidth]{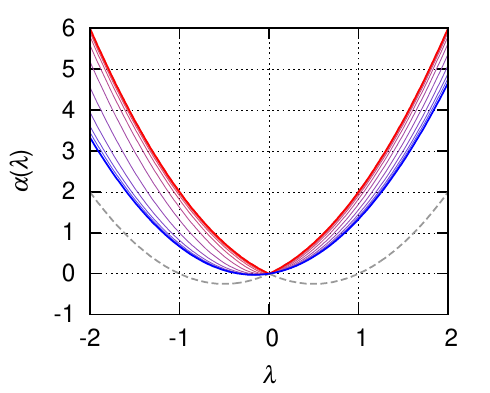}
  \includegraphics[width=0.4\textwidth]{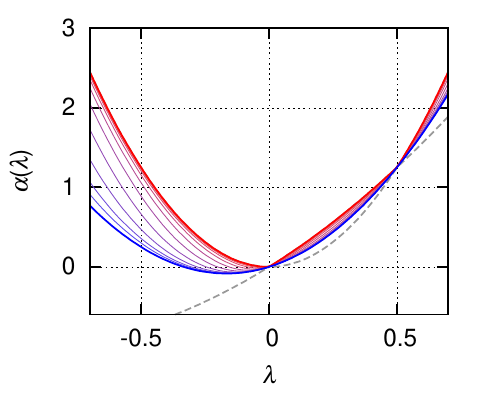}
  \includegraphics[width=0.4\textwidth]{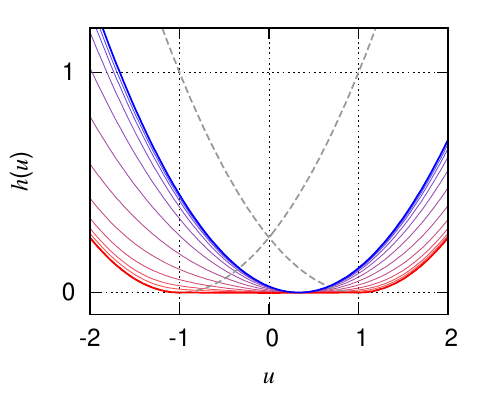}
  \includegraphics[width=0.4\textwidth]{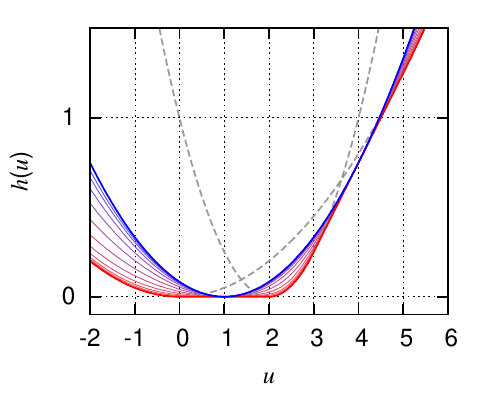}
  \caption{Simple molecular motor models (top row), scaled cumulant generating
    function $\alpha(\lambda)$ (middle row) and corresponding rate functions
    $h(u)$ (bottom row). The left column refers to the motor model with
    dichotomous drift velocity and the right column to the motor model with
    stochastic detaching from the track. The functions are shown as thin
    curves with the internal (dimensionless) timescale $\eta$ ranging from
    0.01 to 10 in 10 steps with logarithmic spacing. The thick blue curves
    show the limit $\eta\to\infty$ and the thick red curves the limit $\eta\to
    0$ in Eq.~\eqref{eq:CGF_zero} with dashed continuations indicating the
    constituting parabolas $\alpha_i^0(\lambda)$ and $h_i(u)$,
    respectively. Parameters for the left column: $w_{12}=1$, $w_{21}=2$,
    $v_{1,2}=\pm 1$, $D_1=1=D_2$. Right column: $w_{12}=1=w_{21}$, $v_1=2$,
    $v_2=0$, $D_1=1$, $D_2=5$. }
  \label{fig:twostate}
\end{figure}

Typical features of the cumulant generating function and the rate function
can be illustrated in a two-state model as a simple example. The two states with drift velocities $v_i\equiv\mu_i f_i$ and diffusion constants $D_i$
are connected via the transition rates $\eta w_{12}$ and $\eta w_{21}$. The
dimensionless parameter $\eta$ is introduced to study the effect of
the time scale of the internal transitions. This system can be used, for
example, to model a single molecular motor walking along a one-dimensional
track, as shown in the top row of Fig.~\ref{fig:twostate}. For motors that can switch between
forward and backward direction, like helicases \cite{mano13} or
complexes of kinesin and dynein \cite{muel08}, the parameters are chosen as
$v_1>0$, $v_2<0$. and $D_1=D_2$. For a unidirectional motor that can detach
from and re-attach to the track \cite{parm04,chou11}, $v_1>0$, $v_2=0$ and
$D_2>D_1$ would be an appropriate choice. The matrix \eqref{eq:L_general_mod}
for these types of two state systems reads
\begin{equation}
  \mathcal{L}(\lambda,\eta)=\left(
    \begin{array}{cc}
      -\eta w_{12}+v_1\lambda+D_1\lambda^2 & \eta w_{21} \\
      \eta w_{12} & -\eta w_{21}+v_2\lambda+D_2\lambda^2    \end{array}
\right).
\end{equation}
Its maximal eigenvalue is, as in general for $2\times 2$-matrices,
\begin{equation}
  \alpha(\lambda,\eta)=\tr\mathcal{L}(\lambda,\eta)/2+\sqrt{(\tr\mathcal{L}(\lambda,\eta))^2/4-\det\mathcal{L}(\lambda,\eta)},
\label{eq:2x2ev}
\end{equation}
where $\tr \mathcal{L}$ and $\det \mathcal{L}$ denote the trace and the
determinant of $\mathcal{L}$, respectively.
This cumulant generating function is plotted in Fig.~\ref{fig:twostate} for
the two simple motor models.  As a universal characteristic, the cumulant
generating function exhibits kink-like features for slow internal dynamics
(small $\eta$) and approaches a smooth parabola for fast internal dynamics
(large $\eta$).  For the model with the dichotomous drift velocity, there is
only one kink in the cumulant generating function around $\lambda=0$. The
cumulant generating function for the model with stochastic detaching exhibits
a second kink at $\lambda\simeq 0.5$.  The corresponding rate functions $h(u)$
in Fig.~\ref{fig:twostate} have been calculated numerically via the Legendre
transformation \eqref{eq:legendre2}. Thereby, the kinks transform to intervals
with constant (or zero) slope.

\subsection{Limiting cases: Slow and fast internal dynamics}
\label{sec:timesep}
The general properties of the cumulant generating function and the rate function can
be understood by considering the limiting cases of slow and fast internal
dynamics. In both cases, the timescales of the internal dynamics and the
translational diffusion process get separated, allowing for an explicit
calculation of the rate function. As in the two simple examples above, we
rewrite the transition rates as $w_{ij}\mapsto\eta w_{ij}$ and
$r_{i}\mapsto\eta r_{i}$, yielding the modified matrix \eqref{eq:L_general}
\begin{equation}
\label{eq:L_general_mod}
  \mathcal{L}_{ij}(\lambda,\eta)=\eta L_{ij}+(\mu_i f_i\lambda+D_i\lambda^2)\delta_{ij}
\end{equation}
with $L_{ij}$ from Eq.~\eqref{eq:markovmatrix}.
The stationary distribution $\pstat_i$ is unaffected by this scaling, hence
the velocity $v$ in Eq.~\eqref{eq:velocity} is independent of $\eta$ as
well. In contrast, the effective diffusion coefficient \eqref{eq:Deff} depends
strongly on $\eta$. 
With an expansion around $\lambda=0$, it can be
calculated exactly as
\begin{equation}
  D_\mathrm{eff}(\eta)=\sum_i\pstat_i D_i+\eta^{-1}\sum_i q_i v_i,
  \label{eq:Deffdiscrete}
\end{equation}
where the vector $q_i$ (the first order correction to the eigenvector of
$\mathcal{L}(\lambda)$) is defined as the unique solution of the linear system
of equations
\begin{equation}
  \sum_j L_{ij} q_j=(v-\mu_i f_i)\pstat_i,\qquad \sum_iq_i=0.
\end{equation}

The case of slow internal dynamics corresponds to $\eta\ll 1$.  In the limit $\eta\to 0$, the eigenvalues of the matrix
\eqref{eq:L_general_mod} are simply given by its diagonal elements
$\alpha_i^0(\lambda)\equiv\mu_if_i\lambda+D_i\lambda^2$. Thus, the limit of the
cumulant generating function is given by
\begin{equation}
  \label{eq:CGF_zero}
  \lim_{\eta\to 0}\alpha(\lambda,\eta)=\max_i\alpha_i^0(\lambda).
\end{equation}
This function can have kinks where different eigenvalues $\alpha_i^0(\lambda)$
intersect. Typically, if not all $\mu_i f_i$ are equal, there is at least one
such kink at $\lambda=0$. In the limit $\eta\to 0$ the cumulant generating
function is actually non-analytic at these kinks. In contrast, for all finite
values $\eta>0$, the Perron-Frobenius theorem ensures that all eigenvalues of
$\mathcal{L}_{ij}(\lambda,\eta)$ are distinct, so that the intersections
vanish and $\alpha(\lambda)$ is still analytic everywhere \cite{bapa97}. The
increasing curvature of $\alpha(\lambda)$ with decreasing $\eta$ is also
reflected in the divergence of the effective diffusion coefficient
$D_\mathrm{eff}(\eta)=\partial_\lambda^2\alpha(\lambda,\eta)|_{\lambda=0}$ in
Eq.~\eqref{eq:Deffdiscrete}.  For non-zero but still small values of $\eta$,
corrections to the eigenvalue entering in the cumulant generating function
$\eqref{eq:CGF_zero}$ can be obtained via perturbation theory as
\begin{equation}
  \alpha_i(\lambda,\eta)=\alpha_i^0(\lambda)-\eta
    r_i+\eta^2\sum_{i\neq j}\frac{w_{ij}w_{ji}}{\alpha_i^0(\lambda)-\alpha_j^0(\lambda)}+\mathcal{O}(\eta^3).
\end{equation}
This approximation works well for most values of $\lambda$, but not in
the region of the kinks, where the eigenvalues become degenerate. 

For the calculation of the rate function it plays a crucial role in which
order the long-time limit in Eq.~\eqref{eq:ldf_def} and the limit $\eta\to 0$
are performed. If the limit $\eta\to 0$ is taken first, the finite time
probability distribution simply reads
\begin{equation}
\fl  p_i(x,t)=\sum_i\int_{-\infty}^{\infty}\mrd x_0\,p_i(x_0,0)(4\pi
  D_it)^{-1/2}
\exp\left[\frac{-(x-x_0-\mu_i f_i t)^2}{4D_it}\right].
\end{equation}
Provided that the initial distribution $p_i(x_0,0)$ is non-zero for all $i$,
the corresponding rate function \eqref{eq:ldf_def} is then
\begin{equation}
  \label{eq:ldf_nonconvex}
  \tilde h(u)=\min_i h_i(u)\equiv \min_i \frac{(u-\mu_i f_i)^2}{4D_i},
\end{equation}
which is typically a non-convex function.  In contrast, if the long-time limit
is taken first, the rate function is uniquely determined by the Legendre
transform \eqref{eq:legendre2} of the analytic function $\alpha(\lambda)$ for
any finite $\eta>0$. The limit $\eta\to 0$ then leads to the convex envelope $h(u)$ of $\tilde h(u)$ in Eq.~\eqref{eq:ldf_nonconvex} as rate function. This order of the limits is the one
relevant for approximating the rate function at small, non-zero values of $\eta$. For every kink in the cumulant generating function $\alpha(\lambda)$ at $\lambda=\lambda_\mathrm{k}$, the rate function $h(u)$ depends linearly on $u$ with the slope $h'(u)=\lambda_\mathrm{k}$ on the interval
\begin{equation}
	\alpha(\lambda_\mathrm{k}^-)<u<\alpha(\lambda_\mathrm{k}^+).
\label{eq:linearinterval}
\end{equation}

For the simple two-state models discussed above the limit \eqref{eq:CGF_zero}
is shown in red in Fig.~\ref{fig:twostate}. In both models there is a kink in
the cumulant generating function at $\lambda=0$. In the model with the
stochastic detaching, the parabolas for the $\eta\to 0$ limit intersect a
second time at $\lambda=v_1/(D_2-D_1)$.  In the Legendre transformed picture
of the rate function, the kink at $\lambda=0$ results in a flat plateau (slope
0) around $u=0$. This plateau forms the convex envelope between the minima of the
parabolas in Eq.~\eqref{eq:ldf_nonconvex} at $u=v_1$ and $u=v_2$, respectively. For the
model with stochastic detaching, there is a further interval with linear slope
$v_1/(D_2-D_1)$ between $u=v_1(D_2+D_1)/(D_2-D_1)$ and
$u=2v_1D_2/(D_2-D_1)$. Within these intervals, fluctuations in the
displacement of the system are dominated by the fluctuations of the internal
variable. Beyond these ranges (especially for $\lambda\to\pm\infty$),
fluctuations are dominated by translational noise and occur almost exclusively
if the internal variable happens to be constant most of the time.

With increasing internal speed $\eta$ of the internal dynamics, the kinks in
the cumulant generating function and the linear regions in the rate function
are softened until both functions approach a parabolic shape shown in blue in
Fig.~\ref{fig:twostate}.  This behavior can be understood by considering the
limit of fast internal dynamics in general. For $\eta\to\infty$, the
eigenvector associated with the largest eigenvalue of the matrix
\eqref{eq:L_general_mod} approaches the stationary distribution
$\pstat_i$. Starting from this distribution, first order perturbation theory
in $1/\eta$ yields the cumulant generating function
\begin{equation}
  \label{eq:gen_func_fast}
  \alpha(\lambda)=v\lambda+\bar D\lambda^2+\mathcal{O}(1/\eta)
\end{equation}
and the rate function
\begin{equation}
  \label{eq:ldf_fast}
h(u)=\frac{(u-v)^2}{4\bar D}+\mathcal{O}(1/\eta),
\end{equation}
which corresponds effectively to the diffusion of a driven passive particle with drift velocity $v$ and diffusion
coefficient $\bar D\equiv\pstat_iD_i$. This diffusion
coefficient is equal to the limiting value $D_\mathrm{eff}(\eta\to\infty)$ in Eq.~\eqref{eq:Deffdiscrete}.

\section{Janus particles}
\label{sec:janus}
\begin{figure}
  \centering
  \includegraphics[width=0.7\textwidth]{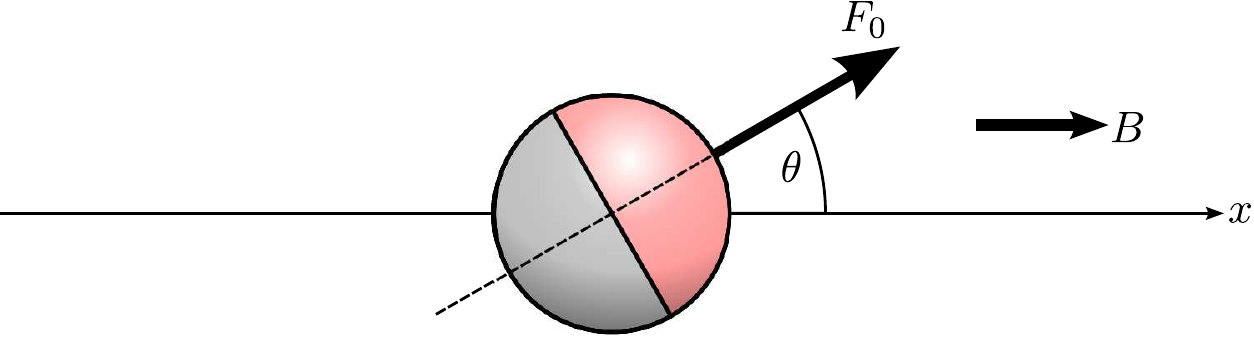}
  \caption{Model for a Janus particle. The internal state is characterized by
    the azimuthal angle $\theta$ between the propulsion force $F_0$ and the
    direction of the coordinate $x$. The propulsion force is aligned with the
    dipole moment, that interacts with the homogeneous field $B$ in
    the $x$-direction.}
  \label{fig:janus_model}
\end{figure}
\subsection{Model}
Janus particles represent another important class of active particles that can
be described within the formalism presented above. As an effective model,
illustrated in Fig.~\ref{fig:janus_model}, we assign to a particle a
propulsion force with constant magnitude $F_0$ acting in the direction of the
instantaneous orientation of the particle. Since the particle is subject to
rotational diffusion, the component of the propulsion force in the direction
of an individual coordinate $x$ varies with time. We assume that both the
rotational and the translational diffusion coefficient are independent of the
orientation, which should be a good approximation for spherically symmetric
particles. In the absence of external potentials, the symmetry of the
configuration leads to a vanishing average velocity in the $x$-direction. Since
the propulsion mechanism drives the system out of equilibrium, breaking this
symmetry leads to a non-vanishing velocity. For example, such a symmetry
breaking could be achieved by an asymmetrically patterned substrate, modeled
as an asymmetric periodic potential \cite{volp11,ao15} or through interaction
with an external field, for example the gravitational field
in gravitaxis \cite{pala10,encu11} or a magnetic field in magnetotaxis \cite{kian15}. A symmetry breaking that can
easily be implemented in the present formalism is a dipolar interaction of the
Janus particle with a homogeneous external field. We identify the coordinate
$x$ with the direction of the field and assume that the dipole moment of the
Janus particle is aligned with the orientation of the propulsion force. The
strength of the interaction, i.e., the dipole moment multiplied by the field
strength in units of $k_\mathrm{B}T$, is denoted as $B$ in the following. We
assume that the direction of the coordinate $x$ is chosen such that $B>0$.

\subsection{Disk-like Janus particles}
First we consider two dimensional, i.e.\ ``disk-like'', Janus
particles, as in Ref.~\cite{hage09} (without field) and, more recently, Ref.~\cite{hanc15}. In this case the angle $\theta$ between the propulsion force and
the direction of the coordinate $x$ fully characterizes the internal state of
the particle, replacing the discrete variable $i$ in
Eq.~\eqref{eq:langevin}. The Langevin equation for the translation of the
Janus particle  reads
\begin{equation}
  \label{eq:langevin_trans}
  \dot x(t)=\mu F_0\cos\theta+\zeta(t)
\end{equation}
with the mobility $\mu$ and correlations of the noise $\zeta(t)$ as in
\eqref{eq:langevin} with a uniform diffusion coefficient $D$. The rotational
motion of the particle is described by the Langevin equation
\begin{equation}
  \label{eq:langevin_rot}
  \dot \theta(t)=-\mu_\mathrm{r}\,k_\mathrm{B} TB \sin\theta+\zeta_\mathrm{r}(t)
\end{equation}
with the rotational mobility $\mu_\mathrm{r}\equiv
D_\mathrm{r}/(k_\mathrm{B}T)$. The rotational noise term obeys
$\mean{\zeta_\mathrm{r}(t)}=0$,
$\mean{\zeta_\mathrm{r}(t)\zeta_\mathrm{r}(t')}=2D_\mathrm{r}\delta(t-t')$,
and is uncorrelated with the translational noise $\zeta(t)$.  The evolution of
the joint probability $p(x,\theta,t)$ is governed by the Fokker-Plank equation
\begin{equation}
  \label{eq:fp_disk}
  \partial_t p(x,\theta,t)=[D_\mathrm{r}L_\mathrm{r}(B)-\mu F_0\cos\theta\,\partial_x+D\partial_x^2]p(x,\theta,t),
\end{equation}
which is a continuous version of Eq.~\eqref{eq:fokkerplanck} with the
transition matrix $L_{ij}$ replaced by the operator
\begin{equation}
  \label{eq:Lrot}
  L_\mathrm{r}(B)\equiv\partial_\theta^2+B\partial_\theta\sin\theta
\end{equation}
for the rotational diffusion of the particle in the field. Analogously, the
cumulant generating function $\alpha(\lambda)$ for the distribution of the
displacement $x$ is given by the maximal eigenvalue of the operator
\begin{equation}
  \label{eq:diskoperator}
  \mathcal{L}(\lambda)=D_\mathrm{r}L_\mathrm{r}(B)+\mu F_0 \cos\theta\,\lambda+D\lambda^2.
\end{equation}
By rescaling to the dimensionless variable 
\begin{equation}
  \scl\equiv\lambda\mu F_0/D_\mathrm{r},  
\end{equation}
this maximal eigenvalue can be written as
\begin{equation}
  \label{eq:full_cgf_janus}
  \alpha(\lambda)=D\lambda^2+D_\mathrm{r}\,a\left(B,\lambda\mu
    F_0/D_\mathrm{r}\right),
\end{equation}
where $a(B,\scl)$ is defined as the maximal value of $a$ for which the
differential equation 
\begin{equation}
  \label{eq:odea} \tilde{\mathcal{L}}(B,\scl,\theta)\psi(\theta)\equiv[L_\mathrm{r}(B)+\scl\cos\theta]\psi(\theta)=a\psi(\theta)
\end{equation}
has a $2\pi$-periodic solution.

Without external field, i.e.\ for $B=0$, this
equation is identical to the Mathieu equation and $a(0,\scl)$ can be
expanded as \cite{abramowitz}
\begin{equation}
  a(0,\scl)=\frac{1}{2}\scl^2-\frac{7}{32}\scl^4+\frac{29}{144}\scl^6+\mathcal{O}(\scl^8).
\end{equation}
Due to the obvious symmetry 
\begin{equation}
	a(0,\scl)=a(0,-\scl),
\label{eq:trivialsymmetry}
\end{equation}
all odd terms vanish in this expansion.
From this expansion one can derive the low order cumulants of the distribution
$p(x,t)$ as
\begin{equation}
  \fl C_2=[D+D_\mathrm{r}(\mu F_0/D_\mathrm{r})^2]\,t,\ 
  C_4=-\frac{21}{4}\,D_\mathrm{r}(\mu F_0/D_\mathrm{r})^4\,t,\ 
  C_6=145\,D_\mathrm{r}(\mu F_0/D_\mathrm{r})^6\,t,
\end{equation}
in accordance with the results for $C_2$ and $C_4$ in Ref.~\cite{hage09}.
It should be noted, however, that this Taylor expansion has a finite radius of convergence beyond which one has to rely on a numerical evaluation of the eigenvalue.

\subsection{A new symmetry}

Remarkably, the trivial symmetry \eqref{eq:trivialsymmetry} persists even in the more interesting case of a non-zero interaction parameter $B$, though with a shifted center of symmetry.
For every eigenfunction $\psi(\theta)$ of $\tilde{\mathcal{L}}(B,\scl)$,
the shifted function $\psi(\theta+\pi)$ is an eigenfunction of the adjoint
operator $\tilde{\mathcal{L}}^\dagger(B,-B-\scl)$, since
\begin{eqnarray}
\tilde{\mathcal{L}}^\dagger(B,-B-\scl,\theta)\psi(\theta+\pi)
 &=[\partial_\theta^2-B\partial_\theta\sin\theta-\scl\cos\theta]\psi(\theta+\pi)\nonumber\\
 &=\tilde{\mathcal{L}}(B,\scl,\theta+\pi)\psi(\theta+\pi).
\label{eq:symmetry_deriv}
\end{eqnarray}
As a consequence, the eigenvalue $a(B,\scl)$ satisfies the symmetry
\begin{equation}
  a(B,\scl)=a(B,-B-\scl).
  \label{eq:symmetry}
\end{equation}
In terms of the cumulant generating function
\eqref{eq:full_cgf_janus} the symmetry \eqref{eq:symmetry} can be expressed as
\begin{equation}
  \label{eq:pseudosymmetry}
  \alpha(\lambda_0+\lambda)=\alpha(\lambda_0-\lambda)+4D\lambda_0\lambda
\end{equation}
with 
\begin{equation}
  \lambda_0\equiv -D_\mathrm{r} B/(2\mu F_0).
\end{equation}
This new symmetry is different from the well-known Gallavotti-Cohen
symmetry \cite{gall95,lebo99}. The latter holds if the observable of interest is
proportional to the entropy production in the medium along individual
trajectories. However, this is not the case for the displacement of Janus
particles. For example, such a particle could move back and forth with
parallel propulsion force for each direction, thus producing entropy without
net displacement. After the Legendre transformation \eqref{eq:legendre2}, the
symmetry \eqref{eq:pseudosymmetry} can be expressed in terms of the rate
function as
\begin{equation}
  h(u_0+u)=2\lambda_0 u+h(u_0-u)
\end{equation}
with
\begin{equation}
  u_0\equiv 2D\lambda_0=-k_\mathrm{B} T D_\mathrm{r} B/F_0.
\end{equation}
For the probability distribution $p(x)$ of the displacement we obtain the
fluctuation relation
\begin{equation}
  \lim_{t\to\infty}\ln\frac{p(u_0 t+\Delta x)}{p(u_0 t-\Delta x)}=-2\lambda_0
  \Delta x.
  \label{eq:pseudoFT}
\end{equation}
In contrast to the well-established fluctuation theorems that are related to
entropy production, this fluctuation relation refers to a center of symmetry
that is not located at $x=0$. Instead, the center of symmetry at $x=u_0t$
moves in backward direction at constant speed.

\subsection{Numerical calculation of the rate function}
\begin{figure}
  \centering
  \includegraphics[width=0.9\textwidth]{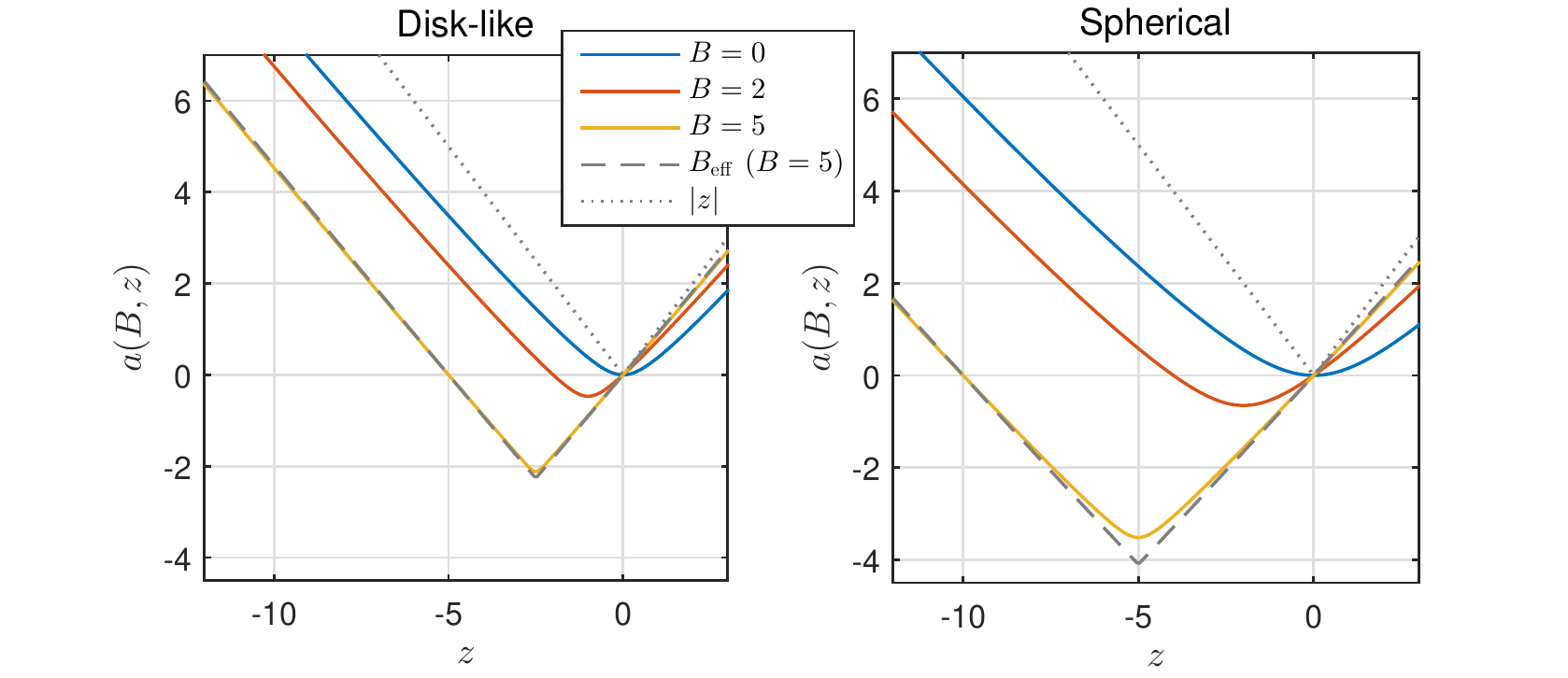}
  \caption{The function $a(B,\scl)$ for a disk-like Janus particle (left) and
    a spherical Janus particle (right) and three different field strengths
    $B$. The gray dashed line shows for $B=5$ the effective field
    approximations \eqref{eq:aapprox} and \eqref{eq:aapprox3D},
    respectively. The gray dotted line shows the obviously insufficient
    approximation $a(B,\scl)=|\scl|$.}
  \label{fig:janusa}
\end{figure}
For an efficient numerical calculation of the eigenvalue $a(B,\scl)$ we expand
the eigenfunction as $\psi(\theta)=c_0+2\sum_{k=1}^\infty c_k\cos(k\theta)$,
which leads to the representation of Eq.~\eqref{eq:odea} as $\sum_{k'}
\tilde{\mathcal{L}}_{kk'}c_{k'}=ac_k$ with the infinite matrix
\begin{equation}
  \label{eq:fourier-matrix}
  \tilde{\mathcal{L}}_{kk'}=-k^2\delta_{k,k'}+(\scl+B k)\delta_{k-1,k'}/2+(\scl-Bk)\delta_{k+1,k'}/2
\end{equation}
(up to the exception $\tilde{\mathcal{L}}_{01}=\scl$). Truncating this matrix
after about 10 to 20 rows and columns already gives very good approximations of
$a(B,\scl)$. The result of this numerical calculation is shown in
Fig.~\ref{fig:janusa} for selected fields $B$. These plots display the
symmetry \eqref{eq:symmetry}. At the center of this symmetry, the function
$a(B,\scl)$ exhibits a kink-like feature, which becomes more pronounced with
increasing $B$. Further away from this kink, the function $a(B,\scl)$ has a
rather small curvature. In Fig.~\ref{fig:ldf_janus}, the numerical result for
$a(B=5,\scl)$ has been used to generate a family of rate functions for
varying rotational diffusion coefficients $D_\mathrm{r}$.

\subsection{Time-scale separation}

The two limiting cases with time scale separation can be analyzed for Janus particles in a similar fashion as for the discrete systems in section \ref{sec:timesep}.
The operator \eqref{eq:diskoperator} is a continuous version of the matrix
\eqref{eq:L_general_mod}, with the rotational diffusion coefficient
$D_\mathrm{r}$ playing the role of the parameter $\eta$. For the extreme cases $D_\mathrm{r}\to 0$ and $D_\mathrm{r}\to \infty$, we can again make use of
the separation of timescales to calculate the cumulant generating function analytically.

\begin{figure}
  \centering
  \includegraphics[width=0.6\textwidth]{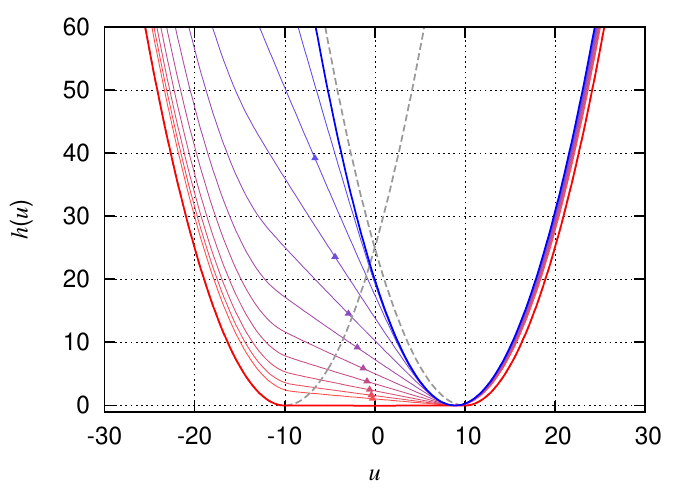}
  \caption{Rate function $h(u)$ for the displacement of a disk-like Janus
    particle with $D=1$, $F_0=10$, and $B=5$. For the thin curves, the
    parameter $D_\mathrm{r}$ ranges from 0.5 to 20 in 10 steps with
    logarithmic spacing. The center of the symmetry \eqref{eq:pseudoFT} is
    marked by triangles. The limit $D_\mathrm{r}\to 0$ is shown as a thick red
    curve with dashed analytic continuations and the limit
    $D_\mathrm{r}\to\infty$ is shown as a thick blue curve. }
  \label{fig:ldf_janus}
\end{figure}

The limit of slow internal dynamics
corresponds to small $D_\mathrm{r}$ and thus large values of $\scl$ in
Eq. \eqref{eq:odea}. The relevant eigenvector of the term $\scl\cos\theta$
for large positive $\scl$ is simply $\delta(\theta)$ with eigenvalue
$\scl$. For large negative values of $\scl$ the relevant eigenvector is
$\delta(\theta-\pi)$ with eigenvalue $-\scl$. Plugging this asymptotic
behavior into Eq.~\eqref{eq:full_cgf_janus} yields the cumulant generating function
\begin{equation}
  \alpha(\lambda)=D\lambda^2+\mu F_0|\lambda|
\end{equation}
in the limit $D_\mathrm{r}\to 0$, which is analogous to
Eq.~\eqref{eq:CGF_zero} in the discrete case. The resulting rate function is
shown in red in Fig.~\ref{fig:ldf_janus}.

Although being asymptotically correct, in the sense that $a(B,\scl)/|\scl|\to
1$ for large $|\scl|$, the approximation $a(B,\scl)\sim|\scl|$ is rather rough for
realistic values of $\scl$. In particular, in Fig.~\ref{fig:janusa}, this
approximation captures neither the shift of the kink nor the slope in the
wings of the function $a(B,\scl)$. The problem is that for these values of $\scl$ the
eigenfunction $\psi(\theta)$ is not yet delta-like but still has a finite
width. Thus, for a better approximation of the eigenvalue for intermediate
values of $\scl$, we choose a representation of the delta-function with finite
width as eigenvector. Since the eigenvector $\psi(\theta)$ for intermediate
$\scl$ somehow interpolates between the known eigenvector $\pstat(\theta)$ at
$\scl=0$ and $\delta(\theta)$ for $\scl\to\infty$, the ansatz
\begin{equation}
\label{eq:psiapprox}
  \tilde\psi(\theta)=\exp[B_\mathrm{eff}(B,\scl)\cos\theta]
\end{equation}
with an effective field $B_\mathrm{eff}(B,\scl)$ that increases with
$\scl$ should be suitable. This parameter can be fixed along with the
approximate eigenvalue $\tilde a(B,\scl)$ by requiring
$[\tilde{\mathcal{L}}(B,\scl,\theta)\tilde\psi(\theta)]/\tilde\psi(\theta)=\tilde
a(B,\scl)+\mathcal{O}(\theta^4)$
for the expansion around the maximum of the eigenvector at $\theta=0$. The
resulting effective field is
\begin{equation}
  B_\mathrm{eff}(B,\scl)=\left[(B-1/2)+\sqrt{(B+1/2)^2+2\scl}\right]/2
\end{equation}
and the approximate eigenvalue becomes
\begin{equation}
  \label{eq:aapprox}
  \tilde a(B,\scl)= \scl+B-B_\mathrm{eff}(B,\scl).
\end{equation}
Although this result is exact only for $\scl=0$ and $\scl\to\infty$, the
numerical results in Fig.~\ref{fig:janusa} show that there is a good agreement with
the actual eigenvalue for all values of $\scl>-B/2$ for strong fields where
the eigenvectors are more concentrated around $\theta=0$. For $\scl<-B/2$, we can make use of the symmetry \eqref{eq:symmetry} to obtain a
global approximation for $a(B,\scl)$.

The limit of fast internal dynamics is obtained for $D_\mathrm{r}\to
\infty$. For an expansion in $z$, we use the stationary distribution
\begin{equation}
  \pstat(\theta)=\frac{\exp(B\cos\theta)}{2\pi I_0(B)}
\end{equation}
to obtain
\begin{equation}
  \label{eq:small_l_janus}
  a(B,\scl)=\int_0^{2\pi}\mrd \theta\,\pstat(\theta)\cos\theta\,\scl+\mathcal{O}(\scl^2)=\frac{I_1(B)}{I_0(B)}\scl+\mathcal{O}(\scl^2),
\end{equation}
where the $I_n(B)$ denotes the modified Bessel functions of the first
kind. Plugging this approximation into Eq.~\eqref{eq:full_cgf_janus} yields
the parabolic approximation of the cumulant generating function analogous to
Eq.~\eqref{eq:gen_func_fast} and the corresponding rate function
\eqref{eq:ldf_fast} with the velocity $v=\mu F_0 I_1(B)/I_0(B)$ and diffusion
coefficient $\bar D=D$. In Fig.~\ref{fig:ldf_janus}, this limiting case is
shown as a blue parabola.

Similar to the case of discrete internal degrees of freedom, the cumulant
generating function \eqref{eq:full_cgf_janus} of a Janus particle exhibits a
kink-like feature. This kink gets smeared out with increasing speed of the
internal dynamics, i.e., with increasing $D_\mathrm{r}$, until the function
approaches the parabola $\alpha(\lambda)=v\lambda+D\lambda^2$. As visible in
the plots of the function $a(B,\scl)$ in Fig.~\ref{fig:janusa}, the location
of the kink coincides with the center of the symmetry
\eqref{eq:pseudosymmetry} at $\lambda=\lambda_0$. As explained for the
discrete model in Sec.~\ref{sec:timesep}, this kink leads in the Legendre
transformed rate function to an interval with constant slope
$\lambda_0$. Based on Eq.~\eqref{eq:linearinterval} and the limits
$-1<\partial_\scl a(B,\scl)<1$, the range where this linear slope occurs can
be limited to $u_0-\mu F_0<u<u_0+\mu F_0$. In this range, the probability
distribution of the displacement does not decay like a Gaussian, as it would
be common for diffusion processes, but rather exponentially. Loosely speaking,
this slower decay in the probability is due to the fact that extreme
fluctuations in this range can be assisted by internal fluctuations of the
rotational degree of freedom.

\subsection{Spherical Janus particles}
For three dimensional, spherical Janus particles we write the orientation in
spherical coordinates $(\theta,\phi)$. 
A distribution over these coordinates is to be understood as a probability per solid angle, hence
integrals over the azimuthal angels $\theta$ require $\sin\theta$ as
Jacobi determinant.
For the joint probability distribution $p(x,\theta,\phi,t)$, the rotational operator in the Fokker-Planck equation \eqref{eq:fp_disk} gets
modified to
\begin{equation}
  L_\mathrm{r}(B)=\frac{1}{\sin\theta}\frac{\partial}{\partial\theta}\left(\sin\theta\frac{\partial}{\partial\theta}\right)
+\frac{1}{\sin^2\theta}\frac{\partial^2}{\partial\phi^2}
+\frac{B}{\sin\theta}\frac{\partial}{\partial\theta}\sin^2\theta,
\end{equation}
Eigenfunctions of the accordingly modified operator $\mathcal{L}(\lambda)$ in
Eq.~\eqref{eq:diskoperator} can be written as products of a $\theta$-dependent
and a $\phi$-dependent part. However, since any dependence on $\phi$ decreases
the corresponding eigenvalue, we can focus on eigenfunctions of the
form $\psi(\theta)$ and thus drop the part of $L_\mathrm{r}(B)$ acting on
$\phi$. The steps leading to the operator $\tilde{\mathcal{L}}(B,\scl,\theta)$
in Eq.~\eqref{eq:odea} then remain unaltered in three dimensions. The adjoint
operator, defined with respect to the natural scalar product on the unit
sphere, can be shown to satisfy the relation
\begin{equation}
 \tilde{\mathcal{L}}^\dagger(B,-2B-\scl,\theta)=\tilde{\mathcal{L}}(B,\scl,\theta+\pi).
\end{equation}
As a consequence, the eigenvalue $a(B,\scl)$ satisfies the symmetry
\begin{equation}
  \label{eq:symmetry3D}
  a(B,\scl)=a(B,-2B-\scl),
\end{equation}
and the property \eqref{eq:pseudosymmetry} of the cumulant generating function still holds, but with a different
center of symmetry $\lambda_0=-D_\mathrm{r}B/(\mu F_0)$.

A numerical calculation of the function $a(B,\scl)$ is possible via an
expansion of the eigenfunctions in Legendre polynomials $P_\ell(\cos\theta)$. The matrix form of
$\tilde{\mathcal{L}}(B,\scl,\theta)$, analogous to \eqref{eq:fourier-matrix},
then reads
\begin{equation}
  \fl\tilde{\mathcal{L}}_{\ell\ell'}=-\ell(\ell+1)\delta_{\ell,\ell'}
+\frac{\ell}{2\ell-1}\left[\scl+B(\ell+1)\right]\delta_{\ell-1,\ell'}
+\frac{\ell+1}{2\ell+3}\left[\scl-B\ell\right]\delta_{\ell+1,\ell'}
\end{equation}
with $\ell,\ell'\geq 0$.
The dominant eigenvalue can be determined numerically in a truncated
approximation of this matrix. The result of the numerics shown in
Fig.~\ref{fig:janusa} reveals that the kink in the cumulant generating function at the
center of symmetry is somewhat less pronounced than for the two-dimensional
model.

For small values of $\scl$, the function $a(B,\scl)$ can be expanded similarly
to Eq.~\eqref{eq:small_l_janus} as
\begin{equation}
  a(B,\scl)=(\coth B-1/B)\scl+\mathcal{O}(\scl^2).
\end{equation}
Accordingly, the average velocity of the Janus particle is given by $v=\mu
F_0(\coth B-1/B)$.
The effective field approximation analogous to Eq.~\eqref{eq:psiapprox} yields for $\scl>-B$
\begin{equation}
  B_\mathrm{eff}=\left[B-1+\sqrt{(B+1)^2+2\scl}\right]/2
\end{equation}
and
\begin{equation}
  \tilde a(B,\scl)=B+\scl+1-\sqrt{(B+1)^2+2\scl},
  \label{eq:aapprox3D}
\end{equation}
which can be continued to $\scl<-B$ via the symmetry \eqref{eq:symmetry3D}.

\section{General model for the active mechanism}
\label{sec:generalization}

In Sec.~\ref{sec:discreteformalism} we have used diffusion with drift as the
arguably simplest possible dynamics for fixed internal state $i$. Molecular
motors that proceed in a stepwise fashion do usually not exert a constant
force as we have assumed in the simple model of Sec.~\ref{sec:casestudy}. An
effective description in terms of the average velocity $v_i=\mu_i f_i$ and
effective diffusion coefficient $D_i$ covers moderate fluctuations quite
well. For extreme fluctuations, however, the approximation of the rate
function for fixed $i$ as a parabola does not capture the previously reported
rich structure of the rate function for molecular motors
\cite{laco08,laco09,piet14} modeled via ratchet models \cite{astu97} or discrete stepping
processes \cite{zimm12}.

\begin{figure}
  \centering
 \raisebox{-\height}{(a)} \raisebox{-\height}{\includegraphics{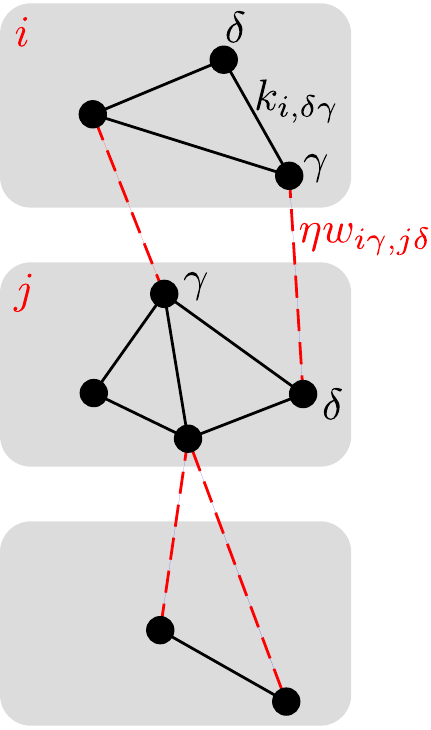}}
 \raisebox{-\height}{(b)} \raisebox{-\height}{
   \begin{minipage}{0.4\linewidth}
     \includegraphics[width=\linewidth]{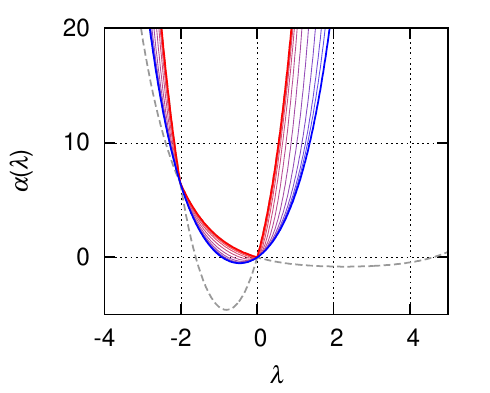}
     \includegraphics[width=\linewidth]{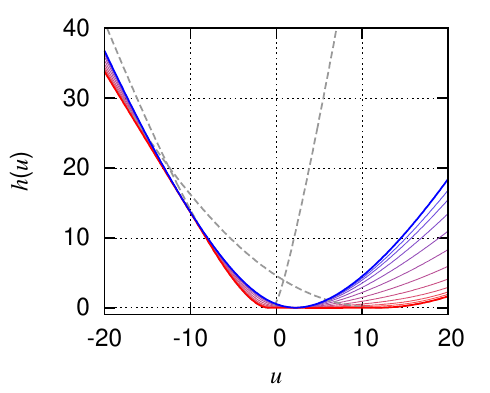}
   \end{minipage}
 }
  \caption{(a) Illustration of the network underlying the generalized
    formalism. Transitions within the mesostates are shown in black,
    transitions between mesostates are shown in red. (b) Cumulant generating
    function (top) and rate function (bottom) for a motor with dichotomous
    motility. The limit $\eta\to 0$ is shown as a thick red curve with dashed
    continuations corresponding to $\alpha_i(\lambda)$ and $h_i(\lambda)$,
    respectively. The limit $\eta\to\infty$ is shown as a thick blue, the thin curves
    correspond to intermediate values of $\eta$ ranging from 0.1 to 100 in 10
    steps with logarithmic spacing. Parameters: $w_{12}=3$, $w_{21}=1$, $k_1^+=15$, $k_1^-=3$, $k_2^+=0.01$, $k_2^-=1.$}
  \label{fig:mesostates}
\end{figure}

In a more detailed and general description, we assign to each of the internal mesostates
(labeled by the Latin letters $i,j,\dots$) further microstates (labeled by
Greek letters $\gamma,\delta,\dots$), as shown in
Fig.~\ref{fig:mesostates}. For molecular motors the mesostates correspond to different motility states, for example, $i$ labels whether
a single motor is attached to a track or encodes the number of attached kinesins and
dyneins in a tug-of-war model. The microstate $\gamma$ denotes the chemical
conformation of the enzyme. 
For microswimmers, the mesostate $i$ labels the swimming direction,
while microstates $\gamma$ label an arbitrarily fine description of the
hydrodynamic flow field around the particle and possibly the geometric
configuration of flagella. In a less detailed, effective description of active
Brownian motion via a velocity dependent force \cite{erdm00,gang13,chau14},
the microstates $\gamma$ correspond to a finely discretized set of velocity states.

We assume Markovian transition rates $k_{i,\delta\gamma}$ from microstate
$\delta$ to microstate $\gamma$ within the mesostate $i$. Associated with this
transition is the displacement $d_{i,\delta\gamma}=-d_{i,\gamma\delta}$ of
the whole system along the coordinate $x$. The transitions between different
mesostates are modeled via the Markovian rates $\eta w_{i\gamma,j\delta}$ from
microstate $\gamma$ of mesostate $i$ to microstate $\delta$ of mesostate
$j$. For simplicity, we assume these transitions to come without
displacement. The dynamics of the system is governed by the master equation
\begin{equation}
  \partial_t
  p(i,\gamma,t)=\sum_{j,\delta}\mathcal{L}_{i\gamma,j\delta}(0)\,p(j,\delta,t)
\label{eq:full_master}
\end{equation}
with
\begin{equation}
  \mathcal{L}_{i\gamma,j\delta}(\lambda)\equiv \eta\Big[
  w_{j\delta,i\gamma}-\delta_{ij}\delta_{\gamma\delta}\sum_{\ell,\varepsilon}
  w_{i\gamma,\ell\varepsilon}\Big]+\delta_{ij}\mathcal{L}_{i,\gamma\delta}(\lambda)
\label{eq:full_master_matrix}
\end{equation}
and
\begin{equation}
  \mathcal{L}_{i,\gamma\delta}(\lambda)\equiv k_{i,\delta\gamma}\mre^{\lambda
    d_{i,\delta\gamma}}-\delta_{\gamma\delta}\sum_{\varepsilon}k_{i,\gamma\varepsilon}.
\end{equation}
The stationary distribution $\pstat(i,\gamma)$ follows upon setting the left
hand side of Eq.~\eqref{eq:full_master} to zero and can be used to calculate
the average velocity
\begin{equation}
  v=\sum_{i,\gamma}\pstat(i,\gamma)\sum_\delta k_{i,\gamma\delta}d_{i,\gamma\delta}.
\end{equation}
The cumulant generating function $\alpha(\lambda)$ for the displacement is
given by the maximal eigenvalue of $\mathcal{L}_{i\gamma,j\delta}(\lambda)$.

In the limit of slow transitions between mesostates, i.e.\ for $\eta\to 0$,
the cumulant generating function is given by
\begin{equation}
  \alpha(\lambda)=\max_i \alpha_i(\lambda),
\end{equation}
where $\alpha_i(\lambda)$ is the maximal eigenvalue of
$\mathcal{L}_{i,\gamma\delta}(\lambda)$.
Kinks in the cumulant generating
function $\alpha(\lambda)$ occur when the two maximal eigenvalues
$\alpha_i(\lambda)$ intersect. In the Legendre transformed rate function
$h(u)$, these kinks result in ranges with linear slope, forming the convex
envelope of the functions $h_i(u)\equiv \max_\lambda[\lambda u-\alpha_i(\lambda)]$.

These features are analogous to what we have described for the simple model
with diffusion and drift in Sec.~\ref{sec:discrete}. Merely the parabolas in
Eqs.~\eqref{eq:CGF_zero} and \eqref{eq:ldf_nonconvex} are replaced by the less
trivial functions $\alpha_i(\lambda)$ and $h_i(u)$, respectively,
corresponding to the dynamics for fixed $i$. The simple model is recovered if
the parabolic approximation of $h_i(u)$ around $u=v$ extends to a wide range
of the parameter $u$, which is the case in the linear response regime where
the driving chemical and/or mechanical forces are weak and for small
step-sizes $d_{i,\gamma\delta}$.

For small non-zero values of $\eta$, the kinks in the cumulant generating
function and the linear slope in the rate function smear out. For larger
$\eta$, the rates $k_{i,\gamma\delta}$ and $\eta w_{i\gamma,j\delta}$ become
comparable. Then, the distinction between meso- and microstates becomes
pointless and one has to diagonalize the full matrix
\eqref{eq:full_master_matrix}. In the limit $\eta\to\infty$, one can simplify
the network by coarse graining microstates that are connected via the
transitions $\eta w_{i\gamma,j\delta}$. Among these, a local
stationary distribution is reached quickly, which can be used to calculate the
effective transition rates in the coarse grained network based on the rates
$k_{i,\gamma\delta}$.

This generalized formalism can be illustrated by reconsidering the motor model with
dichotomous drift velocity in Sec.~\ref{sec:casestudy}. To account for the
discrete stepping of the motor with step length $d$ (e.g., $d=8\,\mathrm{nm}$
for kinesin \cite{coy99}), we here model the motor as a random walker. In the
simplest description, each of the two mesostates $i\in\left\{1,2\right\}$ contains only a single transition
with forward rate $k_i^+$ and backward rate $k_i^-$. These rates usually
follow through coarse graining from a more detailed motor model
\cite{alta12,zimm15}. Switching between the two mesostates occurs at rates
$w_{12}$ and $w_{21}$. The functions relevant for the $\eta\to 0$ limit then read \cite{lebo99}
\begin{equation}
  \alpha_i(\lambda)=k_i^+\mre^{d\lambda}+k_i^-\mre^{-d\lambda}-k_i^+-k_i^+
\end{equation}
and
\begin{equation}
 h_i(u)\equiv u\,\mathrm{arsinh}(a_i u/v_i)-u\,\frac{\mathcal{A}_i}{2}
-\frac{v_i}{a_i}\left[\sqrt{1+\left({a_iu}/{v_i}\right)^2}-\sqrt{1+a_i^2}\right]
\label{eq:ldf_rw}
\end{equation}
with $v_i\equiv k_i^+-k_i^-$, $\mathcal{A}_i\equiv \ln(k_i^+/k_i^-)$, and
$a_i\equiv\sinh(\mathcal{A}_i/2)$.
For $\eta\gg 1$, we can make use of the analytical solution for
the cumulant generating function as the maximal eigenvalue \eqref{eq:2x2ev} of
the matrix
\begin{equation}
  \mathcal{L}(\lambda,\eta)=\left(
    \begin{array}{cc}
      -\eta w_{12}+\alpha_1(\lambda) & \eta w_{21} \\
      \eta w_{12} & -\eta w_{21}+\alpha_2(\lambda)    \end{array}
\right).
\end{equation}
Finally, in the limit $\eta\to\infty$, the cumulant generating function becomes
\begin{equation}
  \alpha(\lambda)=k_\mathrm{eff}^+\mre^{d\lambda}+k_\mathrm{eff}^-\mre^{-d\lambda}-k_\mathrm{eff}^+-k_\mathrm{eff}^+
\end{equation}
with coarse-grained rates 
\begin{equation}
  k_\mathrm{eff}^\pm\equiv(w_{12}k_2^\pm+w_{21}k_1^\pm)/(w_{12}+w_{21})
\end{equation}
and the rate function analogous to Eq.~\eqref{eq:ldf_rw}.
The cumulant generating function and rate function for selected values of the
transition rates are shown in Fig.~\ref{fig:mesostates}.

\section{Summary and perspectives}
\label{sec:summary}

In summary, we have shown that the rate function characterizing extreme
fluctuations of the displacement in active Brownian motion displays a rich
structure, which can be understood in a simple model describing the propulsion
mechanism as a biased diffusion process. While the rate function for passive
Brownian motion is globally parabolic, corresponding to the Gaussian shape of
the distribution, we find for active Brownian motion an interplay between
parabolic and linear sections. This structure of the rate function can be
attributed to the dependence of the propulsion force on the
mesoscopic state of the system. For fluctuations contributing to the parabolic sections, the
internal mechanism typically rests in an individual mesostate, so that the
propulsion force is constant and fluctuations are dominated by translational
diffusion. In contrast, the fluctuations in the linear sections are dominated
by the noise in the internal process, which results in a fluctuating
propulsion force. If the internal process is slow compared to the time scale
of the translational diffusion, the borders between the parabolic and linear
sections of the rate function become sharp. With increasing speed of the
internal process the shape of the rate function gets more smeared out, until
it approaches the globally parabolic rate function of a particle that diffuses
with constant drift and an effective diffusion coefficient.

For experimentally measured distributions, the identification of the different
sections in the rate function could help to elucidate aspects of the hidden internal
states of active systems. For example, fitting of the parabolic sections
would yield information about the propulsion forces $f_i$ and diffusion
coefficients $D_i$ of the corresponding dominant internal states.

For a propulsion mechanism that is more intricate than biased diffusion, like
for stepping motors, we have shown that the general features of the extreme
fluctuations remain the same except that the parabolic sections get replaced
by the more complex rate functions corresponding to the different internal
mesostates.

Although being characteristic for active Brownian
motion, the occurrence of linear sections in the rate function is not
restricted to this type of Brownian motion. For example, a passive but asymmetric
particle pulled with a constant force has different mobilities in different
geometrical alignments and hence also different drift velocities. This fact results in
a similar rate function as we have described for active particles. Similarly,
passive particles in a Poiseuille flow have different drift velocities
depending on the radial position.

For the example of Janus particles with a dipole moment interacting with a
homogeneous field, we have found a new fluctuation relation-like symmetry for
the rate function. Unlike fluctuation relations related to entropy production,
the center of this symmetry is not at velocity zero but at a negative,
field-dependent velocity $u_0$. In experiments, one could make use of this relation
to extract the dipole moment of a Janus particle from the
slope $2\lambda_0$. While this information could in principle be extracted
from a measurement of the average velocity $v$ as well, using the
fluctuation relation would have the advantage of being robust to a spurious
drift from possible external forces. These would only affect the center of the
symmetry $u_0$ but not the slope. This distinction is relevant in experiments
assessing gravitaxis, where gravity and buoyancy jointly cause both a torque and a
translational force.

The fluctuation relation for the Janus particles is complementary to the
fluctuation relation for entropy production also in the respect that it comes
with a linear section of the rate function around the center of symmetry. In
contrast, for the entropy production there is a kink at this center of
symmetry \cite{mehl08,doro11}.

Based on the insight from the discrete model, we expect that the occurrence of
linear sections in the rate function for the displacement is a universal
feature of Janus particles and independent of the geometry and the type of
interaction with external fields. However, the new fluctuation relation we
have proven for the dipolar interaction would not extend to Janus particles
with broken spatial symmetry, as a numerical check for Janus particles in an
asymmetric periodic potential reveals. In future work, it will be a
challenging task to figure out universal conditions under which currents in
arbitrary systems exhibit fluctuation relations with a non-zero center of
symmetry.

\section*{References}

\bibliographystyle{utphys}
\bibliography{refs}

\end{document}